# A Review of Link Aggregation Control Protocol (LACP) as a Link Redundancy in SDN Based Network Using Ryu-Controller

Ali Ibnun Nurhadi[1], Guntur Petrus B.K.[2], Muhammad Firdaus[3], Raditya Muhammad[4]
Department of Electrical Engineering
School of Electrical Engineering and Informatics
Institut Teknologi Bandung, Indonesia
e-mail: ain@students.itb.ac.id[1], guntur_petrus@students.itb.ac.id[2], muhammadfirdaus@students.itb.ac.id[3], radityamuhammad@students.itb.ac.id[4]

*Abstract*— a reliable network is an absolute requirement for telecommunication networks at this time. This is in line with the growing number of users who need a reliable and uninterrupted connection to the server. Link Aggregation is one of the techniques that can be used to provide a reliable network. Link Aggregation System combines two or more links into one logical link, which is characterized by the use of a single IP address. During the communication between host and server, whenever the used link broken or terminated the other active link can act as a redundancy to resume that communication. This mechanism is called link redundancy. This mechanism will not run without the existence of another mechanism that monitors the condition of the links, which is either connected or disconnected. That mechanism is called link-monitoring. Link monitoring used in this simulation is Media Independent Interface (MII).

In this simulation we used Ryu-Controller and mininet Simulator to test the reliability of Link Aggregation-based SDN (Software Defined Network). During the simulation, we analysed failover performance and how the Link Aggregation ditribute the connection accross the links for every which come from different source. From the simulations we confirmed that link redundancy, same data rate for every connection, to some degree, worked as intended.

## I. INTRODUCTION

The number of device users, which are connected to the internet network, are still growing. The emergence of the concept of the Internet of Things and Machine-to-Machine Communication makes the number of users that connect the telecommunications networks are increasingly diverse. In line with this trend, some telecommunications companies estimates the number of devices connected to a telecommunications network by 2020. Cisco predicted that there are 50 Billion devices connected, while IBM predicted 1 trillion devices. An information technology research firm and the firm of United States origin, Gartner, predicted any 6.4 Billion devices beyond smartphones, Tablet PCs, will be connected to the telecommunication network in 2020 [1]. In addition, the Internet, as one of the most popular media, users are growing from day to day. A survey conducted by the Internet Live Stat mentioned that internet users from year to year is always increasing and until the end of the year 2016 at around 3,424,971,237 (3 billion more) [2].

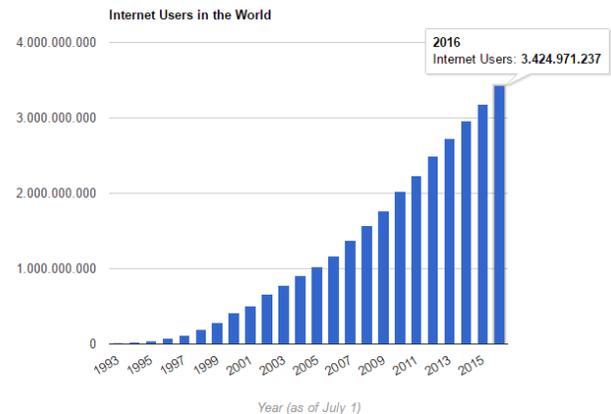

Figure 1. Internet Users in The World

From both the data we can take the conclusion that, as the number of users grew, they need to access higher service. Then the reliability in telecommunication networks, in particularly, the connections against the server must always available whenever and wherever. One way to improve the reliability of the connection towards the server, namely by implementing the mechanism of stackable switches, i.e. stacking multiple switches similar in one place. Stackable switch mechanism able to gives the related benefits of simplicity, scalability, and flexibility network architecture. However, behind some of the profit mechanism of the stackable switch, it also has disadvantages, such as, difficulties in the management switch, a large electrical power needs, performance decreases, the inflexibility and the major drawback are the additional cost for the provision of an additional switch. [3]

As the solution, a system link redundancy that can address the problem of the breakdown in a transmission medium, which resulted in the loss of the network connection, was made. The existence of this link redundancy system, provides redundancy in case the active link is dead or terminated. The communication process will continue, because there is still other active link, which act as a reserve to accommodate the communication process.

System link redundancy in question is a Link Aggregation. Link Aggregation is a system which can combine two or more links into single virtual link. These aggregated links will have one IP address (Internet Protocol) [4].

The organization of this paper is as follow:
1. Part II explain about the supporting theory and classification of the Link Aggregation Control Protocol (LACP) based on the literature on 802.3ad standard and other existing research.
2. Part III discusses system design and simulation scenarios tested.
3. Part IV explained about the analysis of redundancy mechanism of the simulation. In this simulations we saw how Link Aggregation reforwarded an active connections when we terminated the link. We will also see Link Aggregation provide the same data rate to every connection that pass through the link aggregation.

## II. LINK AGGREGATION

The main reference related to the topic (link aggregation) used in this experiments and paper refers to the IEEE 802.3.ad standard. However, on 3 November 2008 [5], LACP protocol is moved from IEEE 802.3.ad standard to 802.1.AX standard, one of the reason is that certain 802.1 layers (such as 802.1X security) were positioned in the protocol stack above Link Aggregation which was defined as an 802.3 sublayer, that noted by David Law in 2006 [6]. So, 802.3ad is the old IEEE working group for what is now known as 802.1AX. LACP is stand for Link Aggregation Control Protocol. Link Aggregation allows one or more links to be aggregated together to form a Link Aggregation Group [7].

### A. Positioning of Link Aggregation within the IEEE 802

Figure 2 describe the position of Link Aggregation within the IEEE 802 architecture.

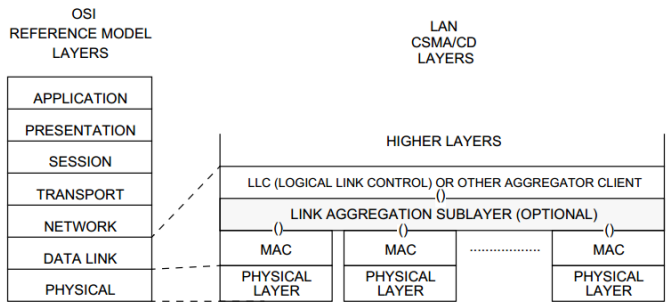

Figure 2. Positioning of Link Aggregation within the IEEE 802 Architecture [7]

### B. Link Aggregation Sublayer Block Diagram

Figure 3 describe the Link Aggregation sublayer block diagram.

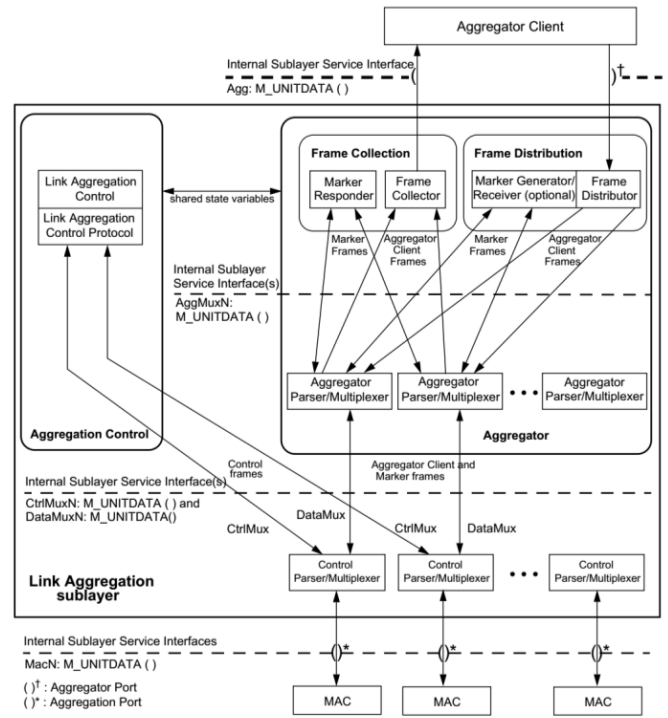

Figure 3. Link Aggregation Sublayer Block Diagram [7]

Figure 3 depicted the Link Aggregation sublayer comprises some functions i.e., Frame Distribution, Frame Collection, Aggregator Parser/Multiplexers, Aggregator, Aggregation Control, and Control Parser/Multiplexers.

### C. Principles of Link Aggregation [7]

Link Aggregation allows an Aggregator Client to treat a set of one or more Aggregation Ports as if it were a single port. In doing so, it employs the following principles and concepts:
1. An Aggregator Client communicates with a set of Aggregation Ports through an Aggregator, which presents a standard ISS interface to the Aggregator Client. The Aggregator binds to one or more Aggregation Ports within a System.
2. It is the responsibility of the Aggregator to distribute frame transmissions from the Aggregator Client to the various Aggregation Ports, and to collect received frames from the Aggregation Ports and pass them to the Aggregator Client transparently.
3. A System may contain multiple Aggregators, serving multiple Aggregator Clients. A given Aggregation Port will bind to (at most) a single Aggregator at any time. An Aggregator Client is served by a single Aggregator at a time.
4. The binding of Aggregation Ports to Aggregators within a System is managed by the Link Aggregation Control function for that System, which is responsible for determining which links may be aggregated, aggregating them, binding the Aggregation Ports within the System to an appropriate Aggregator, and monitoring conditions to determine when a change in aggregation is needed.
5. Such determination and binding may be under manual control through direct manipulation of the state variables of Link Aggregation (e.g., Keys) by a network manager. In addition, automatic determination, configuration, binding, and monitoring may occur through the use of a Link Aggregation Control Protocol (LACP). The LACP uses peer exchanges across the links to determine, on an ongoing basis, the aggregation capability of the various links, and continuously provides the maximum level of aggregation capability achievable between a given pair of Systems.

6. Frame ordering has to be maintained for certain sequences of frame exchanges between Aggregator Clients. The Frame Distributor ensures that all frames of a given conversation are passed to a single Aggregation Port. For any given Aggregation Port, the Frame Collector is required to pass frames to the Aggregator Client in the order that they are received from that Aggregation Port. The Frame Collector is otherwise free to select frames received from the Aggregation Ports in any order. Since there are no means for frames to be misordered on a single link, this guarantees that frame ordering is maintained for any conversation.
7. Conversations may be moved among Aggregation Ports within an aggregation, both for load balancing and to maintain availability in the event of link failures.
8. This standard does not impose any particular distribution algorithm on the Frame Distributor. Whatever algorithm is used should be appropriate for the Aggregator Client being supported.
9. Each Aggregation Port is assigned a MAC address, unique over the Link Aggregation Group and the 802.1Q Bridged LAN (if any) to which the Link Aggregation Group is connected. This MAC address is used as the source address for frame exchanges that are initiated by entities within the Link Aggregation sublayer itself (i.e., LACP and Marker protocol exchanges).
10. Each Aggregator is assigned a MAC address, unique over the Link Aggregation Group and the 802.1Q Bridged LAN (if any) to which the Link Aggregation Group is connected; this address is used as the MAC address of the aggregation from the perspective of the Aggregator Client, both as a source address for transmitted frames and as the destination address for received frames. The MAC address of the Aggregator may be one of the MAC addresses of an Aggregation Port in the associated Link Aggregation Group.

## III. RYU SDN FRAMEWORK

SDN Framework used in this experiment is Ryu SDN Framework. Ryu controller is use Phyton as the programming language. There are two methods used to start the link aggregation function, the static method, in which each network device is instructed directly, and the dynamic method, in which the function is started dynamically using the protocol called Link Aggregation Control Protocol (LACP). Figure 4 depicted the Ryu Standard Architecture.

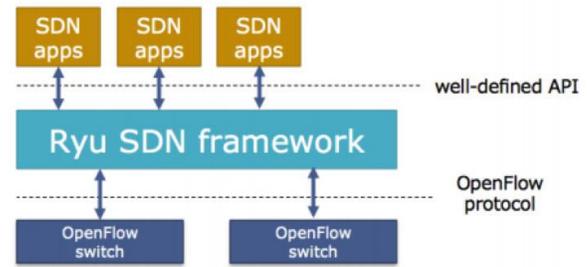

Figure 4. Ryu Standard Architecture

The main guideline and reference of this research refers to RYU SDN Framework, Release 1.0 Book [8]

## IV. SYSTEM DESIGN

Link Aggregation system, built using a client-server network connectivity with the addition of a switch between the client and the server, Link Aggregation is performed by combining two or more links into single virtual link. A virtual link is characterized by the use of a single IP address for multiple links combined (Aggregated links). An aggregator is a mechanism responsible for the distribution of the frame and the collection returned when the communication process takes place. Communication between Aggregated Link using Link Aggregation Control Protocol (LACP) Protocol.

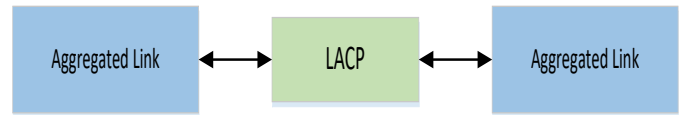

Figure 5. System Block Diagram

Block diagram of this system explains the system in General, that all traffic through the network is set to use LACP Protocol. The existing traffic on this network will be divided on any links that are merged. LACP control the traffic on the network by using a Hash-policy. The use of Hash policy for transmission frame is by calculating the value of the XOR of the MAC address Aggregated Link. In addition, the mechanism set up to frame delivery on the receiver, the frame will be ordered has been received. This system also between a link with the other links that have the same data rate and communication that happens two directions (duplex). to detect the State of the Canal required a link-monitoring so that when one links cut off shipping package will be immediately redirected to use the other links that are still connected.

LACP link aggregation is a dynamic method. That is, the interface from the network device will periodically perform the exchange of data to ensure that communication is possible. In the event of interruptions in data exchange unit LACP, the package will be accepted by the available link. This section will discuss the function of dynamic link aggregation using LACP.

### A. Configuring an Experimental Environment
Scenario experiments conducted aim to analyze the influence of LACP to increased bandwidth on SDN framework using Ryu controller. As a comparison, we create two network topologies with different aggregation number of links as objects of research. The first topology uses 2 link aggregation and the second uses 8 link aggregation.

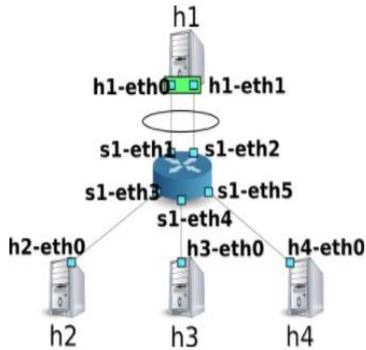

Figure 6. Topology 2 Links

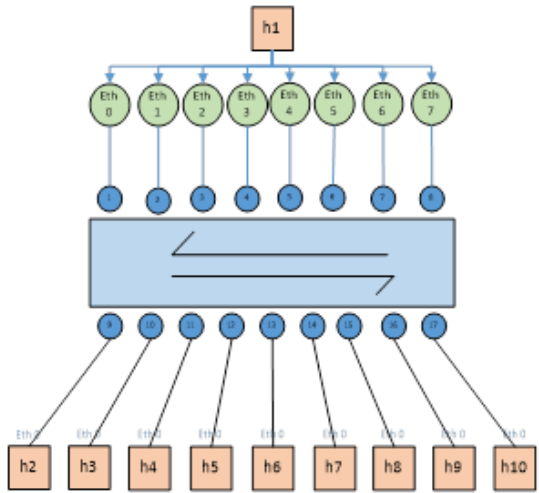

Figure 7. Topology 8 Links

## B. Increasing communication speed

The first step is that we must know the flow entry from switch-1. The result of flow entry from the first scenario uses the topology with 2 links can be seen in Figure 8.

Figure 8. Flow Entry, Topology 2 Links [8]

The result of the topology entry flow by using 8 links can be shown by the Figure 9.

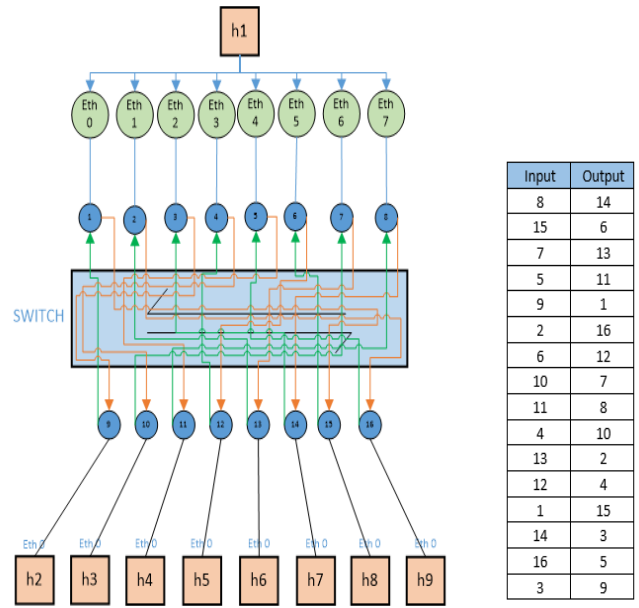

Figure 9. Flow Entry, Topology 8 Links

After knowing the entry flow of each topology, we can measure the bandwidth of the influence of LACP.

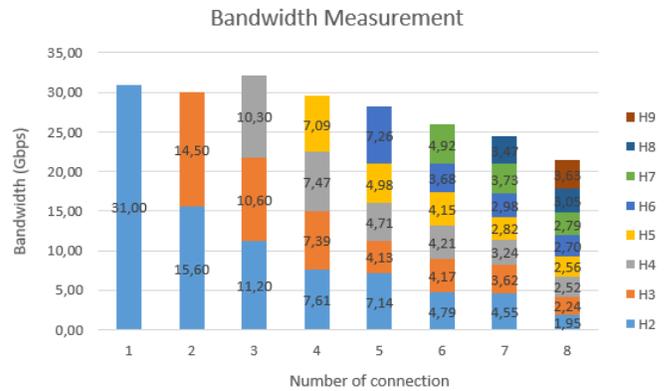

Figure 10. Bandwidth Measurement with 8 Links

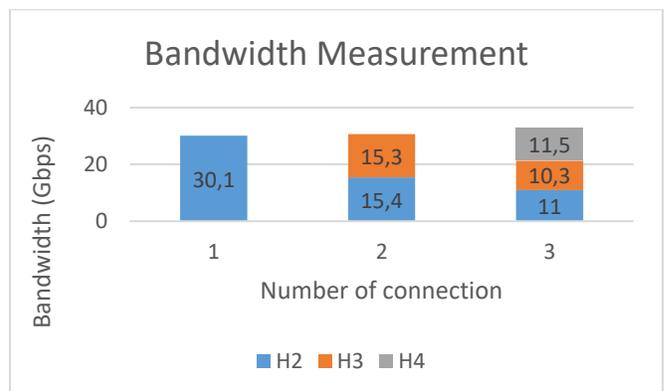

Figure 11. Bandwidth Measurement with 8 Links

## C. Failover

In this part we observe the failover which roughly means how long it takes for an active connections that has its active link down, to be up again. Below we showed the result from 2 link aggregation.

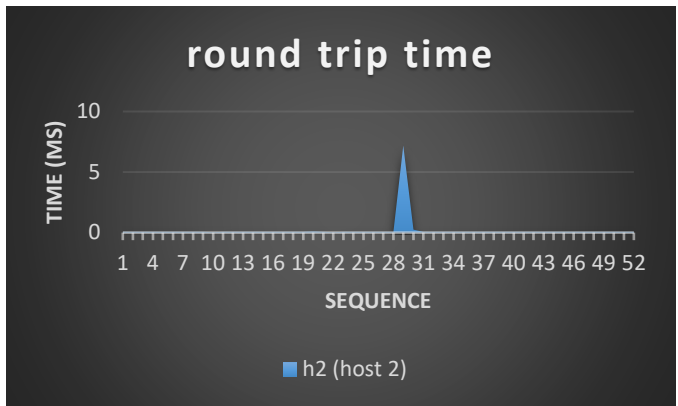

Figure 12. Delay failover at host 2 after link 1 terminated ( 2 link aggregation)

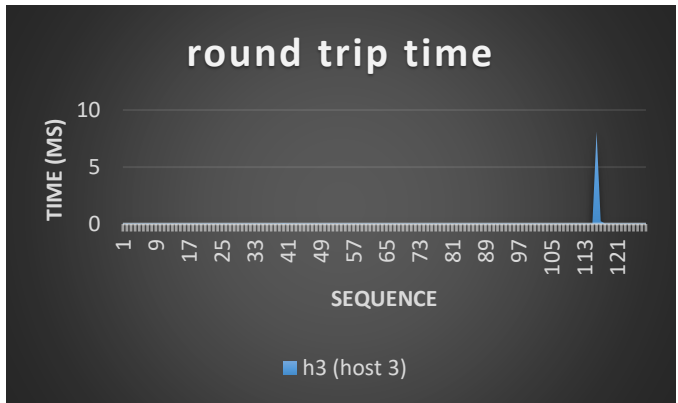

Figure 13. Delay failover at host 3 after link 1 terminated ( 2 link aggregation)

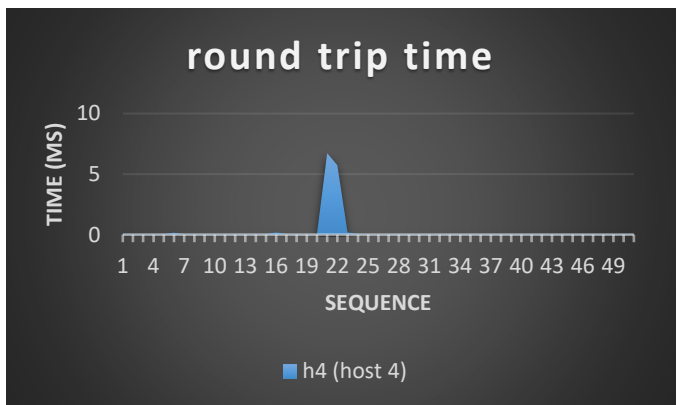

Figure 14. Delay failover at host 4 after link 1 terminated ( 2 link aggregation)

Below we showed the result from 8 link aggregation.

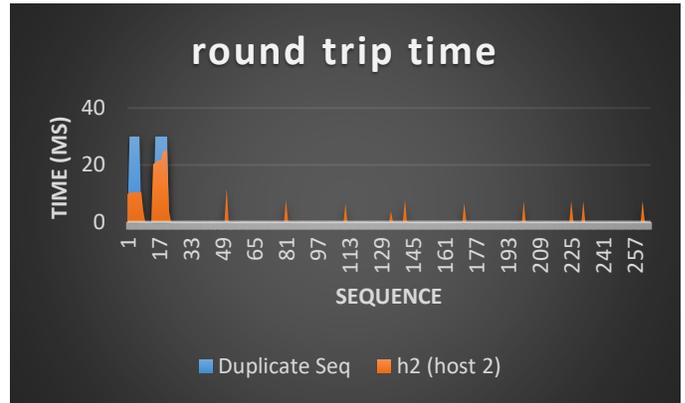

Figure 15. Delay failover at host 2 after link 1 terminated ( 8 link aggregation)

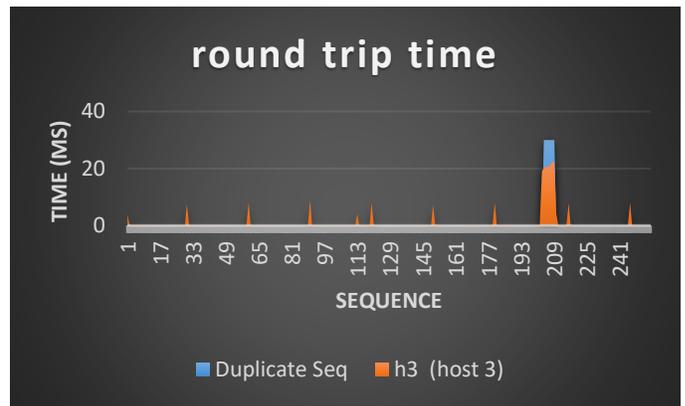

Figure 16. Delay failover at host 3 after link 1 terminated ( 8 link aggregation)

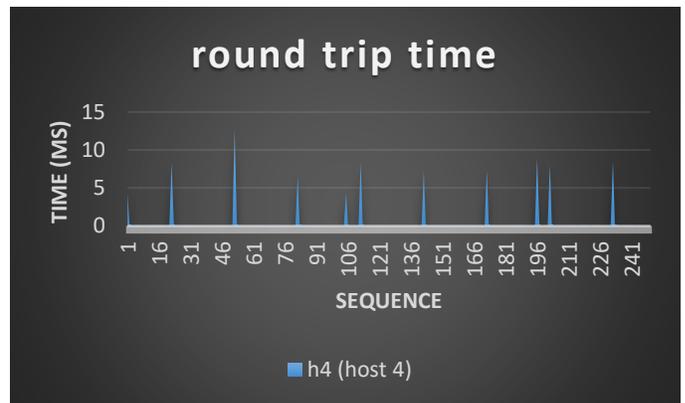

Figure 17. Delay failover at host 4 after link 1 terminated ( 8 link aggregation)

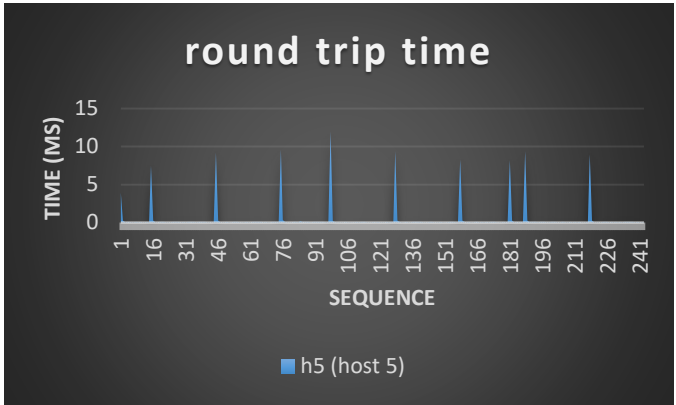

Figure 18. Delay failover at host 5 after link 1 terminated ( 8 link aggregation)

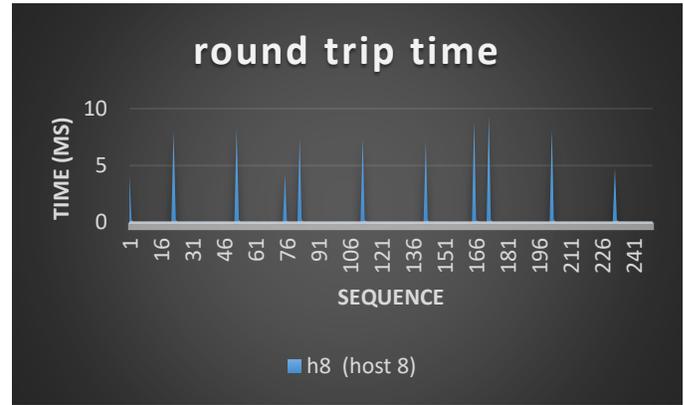

Figure 21. Delay failover at host 8 after link 1 terminated ( 8 link aggregation)

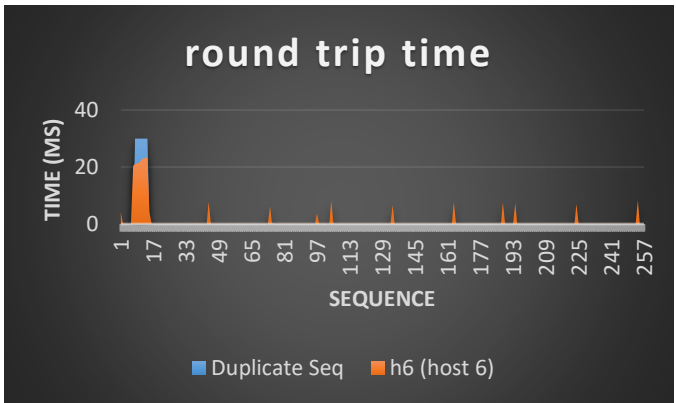

Figure 19. Delay failover at host 6 after link 1 terminated ( 8 link aggregation)

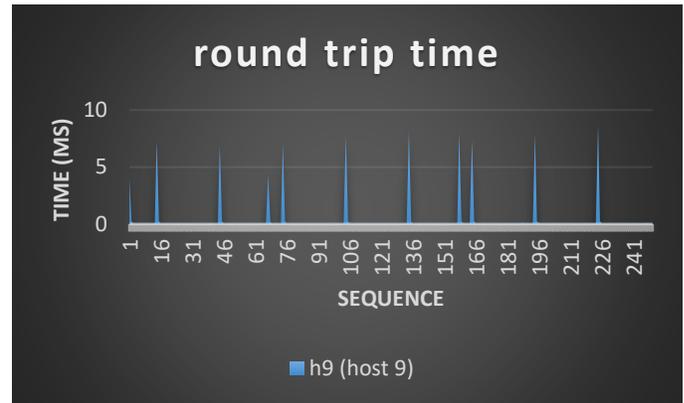

Figure 22. Delay failover at host 9 after link 1 terminated ( 8 link aggregation)

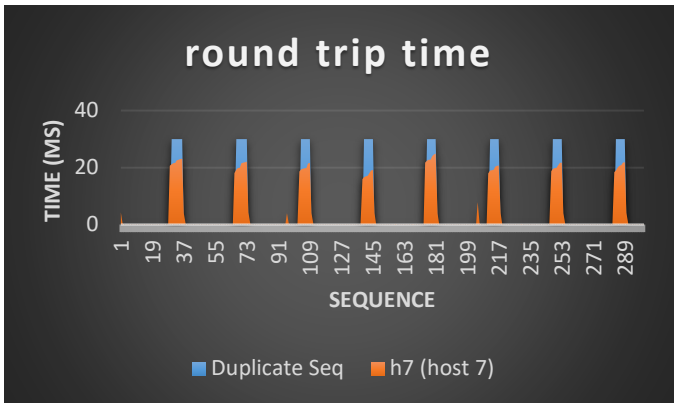

Figure 20. Delay failover at host 7 after link 1 terminated ( 8 link aggregation)

## V. ANALYSES

The LACP protocol give an increasing communication speed for every connection by forwarding each connection to an unused links in link aggregation. As long as the number of simultaneous connections are equal or less than the number of aggregated links, each connection will experience roughly the same bandwidth as the bandwidth of the link (of course provided that the bandwidth bottle neck is the aggregated link). As we saw in our simulation (Figure 8 and 9), both the 2 link scenario and 8 scenario give the expected result. That is, each connection get a its own link. But the bandwidth experienced by each connections didn't go as expected. From Figure 10 and figure 11 we saw that the bandwidth of one single connection is high, but the bandwidth of simultaneous connections is roughly the one connection bandwidth divided by the number of simultaneous connections. This the same for both the 2 scenario and 8 scenario.

The failover for 2 link scenario (figure 12,13, and 14) show an expected result. The burst of delay that we observe in figure (12, 13, and 14) is the result of re-forwarding of connection from host 2 and host 4 (host 2 and host 4 used link 1, which we terminated to observe the failover time). Eventhough the connection of host 3 is undisturbed by the termination of link 1, it felt the impact of link 1 termination when host 2 and host 4 connections are re-forwarded through link 2 (used by host 3).

4For the 8 link aggregation we saw unexpected results. Even though the number of simultaneous connections are equal to the number of links ( eight links) we saw almost periodic burst of delay, another thing that need to be watched is the emergence of duplicate packet that observed by server.

## VI. CONCLUSION

Based on our simulation, Link Aggregation on SDN using Ryu-controller provides :
1. LACP provide roughly equal data rate for every connections and, for every available link, it will be used exclusively by one connection. This remains true as long as the number of simultaneous connections are less than the aggregated links.
2. The bandwidth for every connection that went through each link should be roughly the same with the link bandwidth. This specification is not observed by our experiment. In our experiment The bandwidth is high for one connection and for more than one, the bandwidth divided equally for every simultaneous connections.
3. The failover give disturbance for both the reforwarded connection/s and the undisturbed connection/s as we saw in 2 link aggregation scenario.
4. Ryu-controller is not suitable for 8 link aggregation, because we observe almost periodic burst of delay and duplicate packet is observed by server.

**Reference**
[1] A. Nordrum, "Popular Internet of Things Forecast of 50 Billion Devices by 2020 Is Outdated," *IEEE Spectrum: Technology, Engineering, and Science News*, 18-Aug-2016. [Online]. Available: http://spectrum.ieee.org. [Accessed: 31-Jan-2017]
[2] Number of Internet Users (2016) - Internet Live Stats." [Online]. Available: http://www.internetlivestats.com/internet-users/ #byregion. [Accessed: 30-Nov-2016].
[3] M. Steinbacher and M. Bredel, "LACP Meets OpenFlow – Seamless Link Aggregation to OpenFlow Networks," *TNC15 Conference*.
[4] IEEE Standard Association. 2000." *IEEE Standard for Information Technology - Local and Metropolitan Area Networks - Part 3: Carrier Sense Multiple Access with Collision Detection (CSMA/CD) Access Method and Physical Layer Specifications-Aggregation of Multiple Link Segments*. IEEE Computer Society.
[5] IEEE Standard Association. 2008. *802.1AX-2008 - IEEE Standard for Local and metropolitan area networks--Link Aggregation*. IEEE Computer Society
[6] Law, D. 2006. *"IEEE 802.3 Maintenance"*. Proposal to move Link Aggregation to IEEE 802.1. It is an 802.3 sublayer but it has to go above IEEE Std 802.1x. IEEE Computer Society.
[7] IEEE Standard Association. 2014. IEEE P802.1AX-REV™/D4.54. *Interworking Task Group of IEEE 802.1*. IEEE Computer Society.
[8] RYU project team. 2014. RYU SDN Framework, Release 1.0. RYU project team.